\documentclass[twocolumn,english,showpacs]{revtex4}
\usepackage{babel,amsmath,amssymb,dcolumn}
\usepackage[dvips]{graphics}

\begin{document}

\title{Scalar field evolution in Gauss-Bonnet black holes}

\author{E. Abdalla}
\email{eabdalla@fma.if.usp.br}
\author{R.A. Konoplya}
\email{konoplya@fma.if.usp.br}
\affiliation{Instituto de F\'{\i}sica, Universidade de S\~{a}o Paulo \\
C.P. 66318, 05315-970, S\~{a}o Paulo-SP, Brazil}

\author{C. Molina}
\email{cmolina@usp.br}
\affiliation{Escola de Artes, Ci\^{e}ncias e Humanidades, Universidade de
  S\~{a}o Paulo\\ Av. Arlindo Bettio 1000, CEP 03828-000, S\~{a}o
  Paulo-SP, Brazil}


\begin{abstract}
It is presented a thorough analysis of scalar perturbations in the
background of Gauss-Bonnet, Gauss-Bonnet-de Sitter and
Gauss-Bonnet-anti-de Sitter black hole spacetimes. The 
perturbations are considered both in frequency and time
domain. The dependence of the scalar field evolution  
on the values of the cosmological constant $\Lambda$
and the Gauss-Bonnet coupling $\alpha$ is investigated.
For Gauss-Bonnet and  Gauss-Bonnet-de Sitter black holes, 
at asymptotically late times either power-law or exponential tails
dominate, while for Gauss-Bonnet-anti-de Sitter black hole, the
quasinormal modes govern the scalar field decay at all times. 
The power-law tails at asymptotically late times 
for odd-dimensional Gauss-Bonnet black holes does not depend on $\alpha$,  
even though the black hole metric contains $\alpha$ as a new
parameter. The corrections to quasinormal spectrum due to Gauss-Bonnet
coupling is not small and should not be neglected. For the limit of
near extremal value of the (positive) cosmological constant
and pure de Sitter and anti-de Sitter modes in Gauss-Bonnet gravity we
have found analytical expressions.     
\end{abstract}

\maketitle

%%%%%%%%%%%%%%%%%%%%%%%%%%%%%%%%%%%%%%%%%%%%%%%%%%%%%%%%%%%%%%%%%%%%%%%%%%%%%%%
\section{Introduction}
%%%%%%%%%%%%%%%%%%%%%%%%%%%%%%%%%%%%%%%%%%%%%%%%%%%%%%%%%%%%%%%%%%%%%%%%%%%%%%%

Black holes in more than four spacetime dimensions are of
considerable interest recently  due to the two main reasons: they naturally appear
in string theory, and in extra dimensional brane-world scenarios  \cite{rs}.
According to some of these  scenarios it is possible that
the small higher dimensional black holes can be produced in particles
collisions in Large Hadron Collider. At the same time quantum gravity
may show itself  already at TeV-energy scale. Yet, the  effects of
quantum gravity then may be observed as corrections to classical
General Relativity.     
  
String theory predicts quantum corrections to classical General
Relativity, and the Gauss-Bonnet terms is the first and  dominating
correction among the others. 
Several higher other theories of gravity sustain black hole
solutions. The solution for neutral black hole in Gauss-Bonnet
gravity was obtained by Boulware and Deser \cite{deser} and  
Wheeler \cite{wheeler} . More generally, Lovelock gravity 
\cite{lovelock} has been studied and shown to possess black hole 
solutions with interesting thermodynamical properties
\cite{zanelli,abdcorrea}. 

Thus, the problem of black hole production in transplanckian particle
collisions has attracted considerable interest recently in the context
of large extra dimensions scenarios of TeV-scale gravity. It was
observed that the classical spacetime has large curvature along the
transverse collision plane, as signaled by the curvature invariant
$(R_{ijkl} R^{ijkl}$  and thereby quantum gravity effects, and  higher
curvature corrections to the Einstein gravity, cannot be ignored
\cite{Rychkov}. At the same time we know that  after formation of
such a black hole its evolution has three stages: first, it looses its
``hairs'' coming into Kerr-like phase, then looses angular momentum
transforming to Schwarzschild-like black hole, and finally exerts
strong Hawking evaporation what results in loosing mass (see for
instance \cite{Kanti2005} and references therein). The stage when
black hole perturbations  decay, transforming perturbed black hole
into unperturbed one, is governed by quasinormal modes and is the aim
of our present research.     

In this paper we consider quasinormal perturbations of Gauss-Bonnet
black holes including a non-vanishing cosmological
constant. Quasinormal modes are a very useful tool to uncover
properties of the intrinsic geometry, since the modes characterizes
well the geometry and does not depend on further extrinsic properties,
independent of the geometry itself \cite{Kokkotas-99}. They have been
used successfully in a large class of astrophysical questions, from
black holes to stars. In addition, it has been argued that the
Gauss-Bonnet gravity in asymptotically anti-de Sitter (AdS)
spacetimes may be analyzed through anti-de Sitter/conformal field
theory (AdS/CFT) correspondence within next-to-leading order
\cite{OdintsovJHEP}. In this case the quasinormal modes of the large
Gauss-Bonnet-AdS black hole could find a holographic interpretation
in conformal field theory, as is the cases for the AdS black hole in
Einstein gravity \cite{Horowitz00}.   
   
In \cite{Konoplya05} the quasinormal modes for a charged asymptotically
flat black hole in Gauss-Bonnet gravity were found with the help of
the WKB approach \cite{yer}. The tensor-type gravitational
perturbations for Gauss-Bonnet black hole has been considered recently
in \cite{Dotti}.

To obtain the quasinormal modes we use numerical analysis as well as a
semi-analytical WKB-type treatment. Such an approach is based on the fact
that the wave equation is similar to a Schr\"odinger equation, and
depending on the kind of potential, it makes sense to borrow the methods
used in quantum mechanics in order to define a semi-classical
approximation. This vein has been followed and an approximation for the
quasinormal frequencies has been obtained to a high WKB order \cite{yer}.  
In addition to the frequency domain, we analyze the evolution of 
scalar perturbations in the time domain and find good agreement
between the results found by the two approaches.

We have observed that at asymptotically late
time, power-law tails do not depend on the Gauss-Bonnet coupling and
are the same as for the $d$-dimensional Schwarzschild black hole,
when $d$ is odd. For several simpler particular cases, namely, for
pure de Sitter and anti-de Sitter space-time (without black hole), 
and for extremal Gauss-Bonnet-de Sitter black hole we have found exact 
analytical formulas for (quasi)normal modes. The QNMs for 
Gauss-Bonnet black holes, with coupling $\alpha \sim 1$ predicted by 
string theory, is seemingly different from those of Schwarzschild
black hole. Therefore the GB-corrections to the QN spectrum should not 
be ignored, when considering Tev-scale of quantum gravity scenarios.  
All found modes are damping, what implies stability of Gauss-Bonnet 
black holes against scalar field perturbations.    

The paper is organized as follows: Sec. II represents the
preliminaries of the Gauss-Bonnet-(A)dS metric and its scalar
perturbations. Sec. III is devoted to the methods used in the paper,
namely the WKB method (in the frequency domain), and the
characteristic integration method (in the time domain).
Sec. IV discuss the quasinormal behavior of the Gauss-Bonnet (GB),
Gauss-Bonnet-de Sitter (GBdS) and  Gauss-Bonnet-anti-de Sitter (GBAdS)
black holes. In Sec. V we discuss the future perspective and some
unsolved questions in this field.

%%%%%%%%%%%%%%%%%%%%%%%%%%%%%%%%%%%%%%%%%%%%%%%%%%%%%%%%%%%%%%%%%%%%%%%%%%%%%%%
\section{Gauss-Bonnet Black Hole Solutions}
%%%%%%%%%%%%%%%%%%%%%%%%%%%%%%%%%%%%%%%%%%%%%%%%%%%%%%%%%%%%%%%%%%%%%%%%%%%%%%%

The Einstein-Gauss-Bonnet action in the \mbox{$d$-dimensional} spacetime
model has the form
$$
I = \frac{1} {16 \pi G_{d} } \int d^{d} x \sqrt{-g} R
$$
\begin{equation}
+ \alpha\prime \int d^{d} x \sqrt{-g} (R_{abcd}R^{abcd} - 4 R_{cd}
R^{cd}
+ R^{2} - 2 \Lambda)  
\end{equation}
where $R$ and $\Lambda$ are the $d$-dimensional Ricci scalar and the
cosmological constant, respectively.  The parameter
$\alpha$ represents the (positive) Gauss-Bonnet coupling constant,
which is related to the Regge slope parameter or string scale.    

The Gauss-Bonnet Lagrangian ${\cal L}_{GB}$ is given by
\begin{equation}
{\cal L_{GB}} = R^2 - 4 R_{\mu\nu} R^{\mu\nu}+ 
R_{\mu\nu\rho\sigma} R^{\mu\nu\rho\sigma}\quad .
\label{GB_term}
\end{equation}
One should note that in four dimensions the Gauss-Bonnet term
(\ref{GB_term}) is a total divergency, and yields upon integration 
a topological invariant, namely the genus of the hypersurface
defining the Gauss-Bonnet action (but even in four dimensions there
are interest in the GB correction, as seen  in \cite{olea} for
example).

A metric obtained as a solution of the field equations is given
by
\begin{equation}
ds^{2}=-h(r)dt^{2} + h(r)^{-1}dr^{2}  +r^{2} d\Omega^{2}_{d-2}
\quad ,
\label{metric_form}
\end{equation}
where the function $h(r)$ is given by the expression
\begin{equation}
h(r) = 1 + \frac{r^2}{2\alpha} - \frac{r^2}{2\alpha} 
       \sqrt{1 + \frac{8\alpha\mu}{r^{d-1}} 
       + \frac{8\alpha \Lambda}{(d-1)(d-2)}   } \, .
\label{metric}
\end{equation}
The constant $\mu$ is proportional to the black hole mass and
$d\Omega_{d-2}^{2}$ is the line element of the
($d-2$)-dimensional unit sphere.  The constants 
$\alpha$ and $\alpha\prime$ are connected by the relation:
\begin{equation}
\alpha = 16 \pi G_{d} (d-3) (d-4) \alpha\prime
\end{equation}

We set up a scalar field $\Phi$ on such a background obeying the
Klein-Gordon equation 
\begin{equation}
\Box \Phi = \frac{1}{\sqrt{-g}} \partial_{\mu} (\sqrt{-g} g^{\mu\nu}
\partial_{\nu} \Phi) = 0 \quad .
\end{equation}
In order to separate the wave function in terms of eigenpotential we first
separate the variables as  
\mbox{$\Phi (t,r, \{\theta_{i}\}) = R_{\ell}(t,r)
Y_{\ell m}(\{\theta_{i}\})/r$}. 
As usual we obtain a simple equation for
$R_{\ell}(t,r)$, which is given by the expression
\begin{equation}
4 \frac{\partial^2 R_{\ell}(u,v)}{\partial u\partial v} + 
V(r(u,v)) R_{\ell}(u,v)=0 
\label{simpledalembertian}
\end{equation}
where $u= t-r_\star$, $v=t+r_\star $ and the tortoise coordinate
$r_\star $ is defined by the relation
\begin{equation}
\frac{dr_\star}{dr}=\frac{1}{h(r)}\quad .
\end{equation}
The variables $u$ and $v$ are the light cone coordinates corresponding to the
time and tortoise coordinate. The effective potential for the
scalar field in (\ref{simpledalembertian}) is  
\begin{eqnarray}
V(r) = h(r) \left[ \frac{(d-2)(d-4)}{4 r^{2}} h(r) \right. 
                                  \nonumber \\
                \left. \mbox{} + \frac{d-2}{2r} \frac{d h(r)}{dr}
                  + \frac{\ell (\ell + d - 3)}{r^{2}}  \right] \quad .
\end{eqnarray}

The effective potential is positive definite potential barrier for any
$l$ for GB black hole, and, for $l>0$ for GBdS black hole (For $l=0$
GbdS case, the negative pitch appears). For GBAdS case the potential
diverges at infinity.

%%%%%%%%%%%%%%%%%%%%%%%%%%%%%%%%%%%%%%%%%%%%%%%%%%%%%%%%%%%%%%%%%%%%%%%%%%%%%%%
\section{Numerical and Semi-analytical approaches}
%%%%%%%%%%%%%%%%%%%%%%%%%%%%%%%%%%%%%%%%%%%%%%%%%%%%%%%%%%%%%%%%%%%%%%%%%%%%%%%

\subsection{Characteristic integration}

In \cite{Gundlach-94} a simple but very efficient way of dealing with 
two-dimensional d'Alembertians  has been set up. Along the general lines
of the pioneering work \cite{Price-72},  light-cone variables have been
introduced, leading to (\ref{simpledalembertian}). 

In the characteristic initial value problem, initial data are specified
on the two null surfaces $u = u_{0}$ and $v = v_{0}$. The basic
aspects of the field decay are independent of the initial conditions
(as confirmed by simulations), so we use Gaussian initial conditions.

Since we do not have analytic solutions to the
time-dependent wave equation with the effective potentials introduced,
one approach is to discretize the equation (\ref{simpledalembertian}), and
then implement a finite differencing scheme to solve it
numerically. One possible discretization, used for example in
\cite{Wang-01,Brady-97,Brady-99},  is  
\begin{eqnarray}
\lefteqn{R_{\ell}(N) = R_{\ell}(W) + R_{\ell}(E) -
R_{\ell}(S) }  \nonumber \\  
& & \mbox{} - \Delta^2 V(S) \frac{ R_{\ell}(W) + R_{\ell}(E)}{8} \ , 
\label{discrete1}
\end{eqnarray}
where we have used the definitions for the points: $N = (u + \Delta, v
+ \Delta)$, $W = (u + \Delta, v)$, $E = (u, v + \Delta)$ and $S =
(u,v)$.
Another possible scheme is
\begin{gather}
\left[ 1 - \frac{\Delta^{2}}{16}V(S)\right] R_{\ell}(N) = 
R_{\ell}(E)+ R_{\ell}(W) - R_{\ell}(S)\nonumber \\
-\frac{\Delta^{2}}{16}
       \left[V(S)R_{\ell}(S)+V(E)R_{\ell}(E)+V(W) R_{\ell}(W)\right] \,\,.   
\label{discrete2}
\end{gather}   
Although the second discretization (\ref{discrete2}) is more time consuming
than (\ref{discrete1}), it was observed in \cite{Wang-04}  that
(\ref{discrete2}) is more stable for fields in asymptotically AdS
geometries. With the use of expression  (\ref{discrete1}) or
(\ref{discrete2}), the basic algorithm will cover the  region of
interest in the $u-v$ plane, using the value of the field at three
points  in order to calculate  it at a forth one.  After the
integration is completed, the values of $R_{\ell}$ in the regions of
interest are extracted.

\subsection{WKB analysis}

Considering the Laplace transformation of the
Eq. (\ref{simpledalembertian}) (in terms of $t$ and $r_{\star}$), one
gets the ordinary differential equation     
\begin{equation}
\frac{d^2 \psi_{\ell} (r_\star)}{d r_\star^2} - \left[s^2 + V(r(r_\star))
  \right]\psi_{\ell}(r_\star)=0 \quad .  
\end{equation}
One finds that there is a discrete set of possible values of $s$
such that the function $\psi_{\ell}(r_\star)$ satisfies both
boundary conditions:
\begin{equation}
\lim_{r_{\star} \rightarrow -\infty}\psi_{\ell} \, e^{s r_{\star}}=1 \ ,
\end{equation}
\begin{equation}
\lim_{r_{\star}\rightarrow +\infty}\psi_{\ell} \, e^{-s r_{\star}}=1 \ .
\end{equation}
By making the formal replacement $s=i\omega$, we have the usual quasinormal
mode boundary conditions. The frequencies $\omega$ (or $s$) are
the quasinormal frequencies.

The semi-analytic approach used in this work \cite{yer}
is a very efficient algorithm to calculate the quasinormal
frequencies, which have been applied in a variety of situations
\cite{WKB-ap}.

Under the choice of the positive sign of the real part of
$\omega$, QNMs of Gauss-Bonnet and Gauss-Bonnet-de Sitter black holes
satisfy the following boundary conditions 
\begin{equation}
\psi (r_{\star}) \approx C_\pm \exp( \pm i \omega r_{\star}) \quad
\textrm{as} \quad r_{\star} \rightarrow \pm \infty \quad ,
\end{equation}
corresponding to purely in-going waves at the event horizon and
purely out-going waves at null infinity (or cosmological horizon, if
$\Lambda>0$). For the Gauss-Bonnet-anti-de Sitter geometries, the
effective potential is divergent at spatial infinity (which
corresponds to a finite value of $r_{\star}$, here taken as 0). In the
present work, we assume Dirichlet boundary conditions, setting
$\psi_{\ell} (r_\star = 0)=0$.   

To find the quasinormal modes of the black hole whose  effective
potential has the form of a potential barrier (GB and GBdS black holes)
one can use a high order WKB approach, finding   
\begin{equation}
i \frac{\omega^2 - V_0}{\sqrt{-2V_0^{\prime\prime}}} - L_2 - L_3 - L_4 -
L_5 - L_6 = n + \frac{1}{2} \quad ,
\end{equation}
where $V_0$ is the height and $V_0^{\prime\prime}$ is the second
derivative with respect to the tortoise coordinate of the potential at
the maximum. $L_2$, $L_3$  $L_4$, $L_5$ and $L_6$ are presented in
\cite{yer}. Thus we are able to use this formula for finding the
quasinormal modes of Gauss-Bonnet and Gauss-Bonnet de Sitter black
holes. Yet, for Gauss-Bonnet anti-de Sitter it cannot be applied as
the corresponding potential is divergent at spatial infinity.

Accuracy of WKB approach may be bad for some cases of higher
dimensional black holes. 
We think that it is mainly not because of second  small peak in higher
dimensional case \cite{konoplya03}, \cite{A}: the WKB inaccuracy is
limited by the case $\ell=n$ or $\ell<n$. To judge about accuracy of
WKB method one has to compare the WKB results with results obtained by
an accurate Frobenius procedure. This was done for a $d$-dimensional Schwarzschild
black hole in a paper \cite{B}, where it was shown that for low
overtones ($\ell>n$) the difference between 6th order WKB  and
Frobenius method results  is less then one per-cent. We believe this
signifies the relialability of WKB formulas for $\ell>n$ modes, even for
higher dimensional black holes. After all, for $l>0$ modes, and for
scalar field perturbations considered in this paper, there is no
negative pitch in the potential.

%%%%%%%%%%%%%%%%%%%%%%%%%%%%%%%%%%%%%%%%%%%%%%%%%%%%%%%%%%%%%%%%%%%%%%%%%%%%%%%
\section{Evolution of perturbations: time and frequency domain}
%%%%%%%%%%%%%%%%%%%%%%%%%%%%%%%%%%%%%%%%%%%%%%%%%%%%%%%%%%%%%%%%%%%%%%%%%%%%%%%

In this section we shall discuss the quasinormal and late-time
behavior for scalar field perturbations in the exterior of
Gauss-Bonnet black holes, generally, with a null, positive and
negative $\Lambda$-term, and therefore one has to consider the
correlation of the scalar field evolution with ``global''  parameters:
GB-coupling $\alpha$, $\Lambda$-term, spacetime dimensionality, and
``local'' parameters such as black hole mass $\mu$ and  multipole
number $\ell$.

%%%%%%%%%%%%%%%%%%%%%%%%%%%%%%%%%%%%%%%%%%%%%%%%%%%%%%%%%%%%
\begin{figure}
\resizebox{1\linewidth}{!}{\includegraphics*{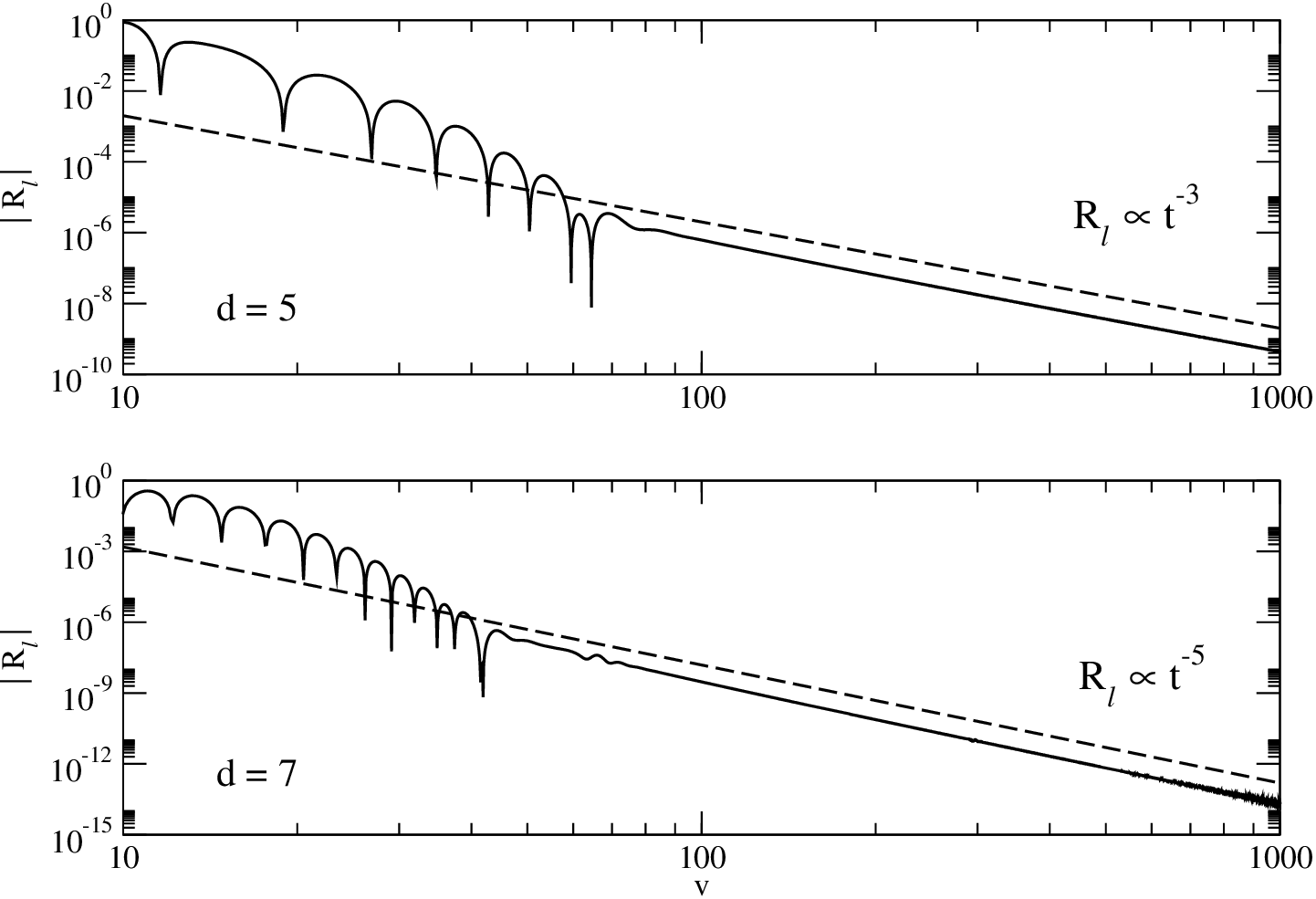}}
\caption{Field decay in the Gauss-Bonnet
    black holes, for \mbox{$d=5$} and $d=7$. It is observed a quasinormal mode
    dominated region. Asymptotically, the field decays as a power-law
    tail (dashed lines).  The parameters in this graph are $\alpha=1$,
    $\mu=1.0$ and $\ell=0$.}
\label{fig_GB1}
\end{figure}
%%%%%%%%%%%%%%%%%%%%%%%%%%%%%%%%%%%%%%%%%%%%%%%%%%%%%%%%%%%%

%%%%%%%%%%%%%%%%%%%%%%%%%%%%%%%%%%%%%%%%%%%%%%%%%%%%%%%%%%%%
\begin{figure}
\resizebox{1\linewidth}{!}{\includegraphics*{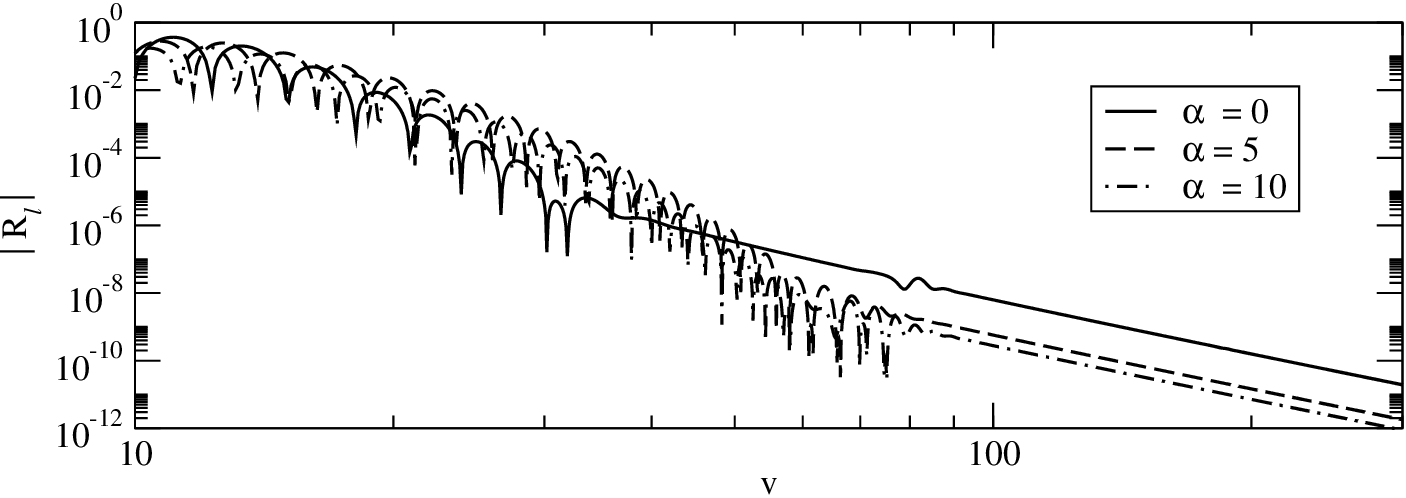}}
\caption{Power-law tails in the Gauss-Bonnet
    black holes. The estimated power-law
    are $R_{\ell} \propto t^{-5.076}$,  $R_{\ell} \propto t^{-5.082}$ and
    $R_{\ell} \propto t^{-5.068}$ for $\alpha = 0$,  $\alpha = 5$ and
    $\alpha = 10$ respectively. The predicted power for $\alpha=0$ is
    $-5$. The parameters in this graph are $d=7$, $\mu=1.0$ and
    $\ell=0$.}   
\label{fig_GB2}
\end{figure}
%%%%%%%%%%%%%%%%%%%%%%%%%%%%%%%%%%%%%%%%%%%%%%%%%%%%%%%%%%%%

\subsection{Gauss-Bonnet black holes}

As seen in a previous work \cite{Konoplya05}, the WKB method allows
very accurate calculations of the quasinormal modes associated with
the field evolution. A complementary analysis can be performed within
the time-dependent picture. For this purpose, we use here a
characteristic initial value algorithm.   

The scenario presented by the WKB calculations is consistent with the
results obtained with time evolution approach. From the wave-functions
calculated with  the characteristic integration routine, it is
observed that, after an initial transient phase, the decay is
dominated by the quasinormal mode ringing. It is possible
to estimate with high precision the oscillatory and exponential decay
parameters using a non-linear fitting based in a $\chi^2$
analysis. We emphasize that the numerical concordance is excellent, as
seen in Tables \ref{GB_QNM_1} - \ref{GB_QNM_4}. 
The results, as compared between WKB approximation and characteristic 
integration agree to an accuracy within a few per-cents for $l>n$ case. 
This small difference must exist, because we compare the data for the
fundamental overtones in frequency domain with time domain data 
where the contribution from all overtones is taken into consideration. 
Unfortunately the WKB accuracy for $l=n=0$ is not satisfactory what
results in large difference between frequency and time domain data for
that case.  

Strictly speaking, the WKB technique we used here converges only
asymptotically. Practically, WKB formula shows good convergence within
several few orders after eikonal approximation. Yet,  
the worse convergence of the WKB method takes place when we deal with 
the intermediate values of $\alpha \approx 1$. That is why, in this
regime, the agreement between the WKB and the characteristic
integration results is the worst.

The imaginary part of the frequency does not
show too much dependence on $\alpha$, yet slightly decrease when
$\alpha$ is increasing. On the other hand, the real part
increases with $\alpha$, though not significantly either. This might be
showing that a quasinormal mode is much more an effect connected with the
local geometry containing the black hole rather than with the global
effect of the geometry, namely, the effect of the existence of an event
horizon matters much more than a detailed dependence on the parameter
$\alpha$. Yet, large enough values of $\alpha$ certainly affect the
quasinormal spectrum: the QNMs are proportional to $\alpha$ in the
regime of large $\alpha$ \cite{Konoplya05}. As $\alpha$ approaches
zero, the QNMs go to those of ordinary $d$-dimensional Schwarzschild 
black hole described by the Tangherlini metric.

However, it does not mean that GB corrections are negligible.
On the contrary, according to  string theory 
the Gauss-Bonnet coupling $\alpha$ should be around $1$. Let us compare 
the results, for instance, for $\ell=2$, $n=0$ QNMs for Schwarzschild and
$\alpha = 1$ GB black holes for $d=6$: for Schwarzschild we have 
$\omega = 1.5965 - 0.3967 i$ (6th order WKB),  for $\alpha =1$ (6th
order WKB) we get $\omega = 1.69654-0.31929 i$ (Note that for this
case convergence is good and the 3th order WKB value is not much
different  $1.67624 - 0.323698 i$). Thus the effect of GB coupling is
about $6.3\%$ in real and more then $20\%$ in the imaginary
part here. For larger values of $\alpha$ it is certainly larger. We
have the same order of difference for other values of $n$ and $\ell$.

\begin{table}
\caption{Values for the quasinormal frequencies for the fundamental
  mode in the Gauss-Bonnet geometry, obtained from sixth order WKB
  method and directly from characteristic data, for
  $d = 5$, $\ell=0,1$ and several values of $\alpha$.} 
\label{GB_QNM_1} 
\begin{ruledtabular}
\begin{tabular}{cccccc}
\multicolumn{2}{l}{ d = 5 }         &
\multicolumn{2}{c}{WKB}&
\multicolumn{2}{c}{Characteristic Integration}\\ 
$\ell$ & $\alpha$ & Re($\omega_0$) & -Im($\omega_0$)  &
Re($\omega_0$) & -Im($\omega_0$)
\\
\hline
\\
0 & 0.1 & 0.389935 & 0.256159 &  0.379 & 0.282 \\
0 & 0.2 &  0.396034 & 0.250548 &  0.383 & 0.272  \\ 
0 & 0.5 &  0.429741 & 0.208293 &  0.391 & 0.246 \\ \\
1 & 0.1 &  0.720423 & 0.255506 &  0.7234 & 0.250  \\
1 & 0.2 &  0.723177 & 0.252684 &  0.7284 & 0.245  \\
1 & 0.5 &  0.739205 & 0.236885 &  0.7443 & 0.228 \\
\end{tabular} 
\end{ruledtabular}
\end{table}

\begin{table}
\caption{Values for the quasinormal frequencies for the fundamental
  mode in the Gauss-Bonnet geometry, obtained from sixth order WKB
  method and directly from characteristic data, for
  $d = 6$, $\ell=0,1$ and several values of $\alpha$.} 
\label{GB_QNM_2}
\begin{ruledtabular}
\begin{tabular}{cccccc}
\multicolumn{2}{l}{ d = 6 }         &
\multicolumn{2}{c}{WKB}&
\multicolumn{2}{c}{Characteristic Integration}\\ 
$\ell$ & $\alpha$ & Re($\omega_0$) & -Im($\omega_0$)  &
Re($\omega_0$) & -Im($\omega_0$)
\\
\hline
\\
0 & 0.1 & 0.735854 & 0.402416 & 0.7109 & 0.412  \\
0 & 0.2 & 0.748053 & 0.391049 & 0.7144 & 0.402 \\
0 & 0.5 & 0.83530  & 0.304837 & 0.7225 &  0.375  \\
0 & 5   & 0.906661 & 0.144820 & 0.9526 & 0.226 \\
0 & 10  & 1.513624 & 0.456935 & 1.5182 & 0.423 \\
0 & 20  & 2.961675 & 0.927128 & 2.878  & 0.952 \\ \\
1 & 0.1 & 1.139007 & 0.415034 & 1.153  & 0.395  \\
1 & 0.2 & 1.136193 & 0.415706 & 1.159  & 0.386  \\
1 & 0.5 & 1.158635 & 0.391339 & 1.176  & 0.363  \\
1 & 5   & 1.790103 & 0.258382 & 1.791  & 0.253  \\
1 & 10  & 3.260415 & 0.498426 & 3.253  & 0.504  \\
1 & 20  & 6.437139 & 0.991374 &  -     &   -    \\
\end{tabular} 
\end{ruledtabular}
\end{table}

\begin{table}
\caption{Values for the quasinormal frequencies for the fundamental
  mode in the Gauss-Bonnet geometry, obtained from sixth order WKB
  method and directly from characteristic data, for
  $d = 7$, $\ell=0,1$ and several values of $\alpha$.} 
\label{GB_QNM_3}
\begin{ruledtabular}
\begin{tabular}{cccccc}
\multicolumn{2}{l}{ d = 7 }         &
\multicolumn{2}{c}{WKB}&
\multicolumn{2}{c}{Characteristic Integration}\\ 
$\ell$ & $\alpha$ & Re($\omega_0$) & -Im($\omega_0$)  &
Re($\omega_0$) & -Im($\omega_0$)
\\
\hline
\\
0 & 0.1 & 1.11738 & 0.546056 & 1.092 & 0.532 \\
0 & 0.2 & 1.13699 & 0.529543 & 1.092 & 0.520 \\
0 & 0.5 & 1.29469 & 0.395111 & 1.092 &  0.493 \\
0 & 5   & 1.39823 & 0.574472 & 1.275 & 0.351 \\
0 & 10  & 1.48053 & 0.385616 & 1.515 & 0.358 \\
0 & 20  & 2.00368 & 0.56458  & 2.001 & 0.567 \\ \\
1 & 0.1 & 1.54573 & 0.577608 & 1.587 & 0.527 \\
1 & 0.2 & 1.53194 & 0.58701 & 1.530 & 0.517 \\
1 & 0.5 & 1.56277 & 0.554551 & 1.609 &  0.489 \\
1 & 5   & 2.01379 & 0.308316 & 1.982 &  0.337 \\
1 & 10  & 2.47824 & 0.243737 & 2.475 &  0.423 \\
1 & 20  & 3.39094 & 0.59452 & 3.387 &  0.597 \\
\end{tabular}
\end{ruledtabular}
\end{table}

\begin{table}
\caption{Values for the quasinormal frequencies for the fundamental
  mode in the Gauss-Bonnet geometry, obtained from sixth order WKB
  method and directly from characteristic data, for
  $d = 8$, $\ell=0,1$ and several values of $\alpha$.} 
\label{GB_QNM_4}
\begin{ruledtabular}
\begin{tabular}{cccccc}
\multicolumn{2}{l}{ d = 8 }         &
\multicolumn{2}{c}{WKB}&
\multicolumn{2}{c}{Characteristic Integration}\\ 
$\ell$ & $\alpha$ & Re($\omega_0$) & -Im($\omega_0$)  &
Re($\omega_0$) & -Im($\omega_0$)
\\
\hline
\\
0 & 0.1 & 1.51702 & 0.694245 & 1.461 & 0.676 \\
0 & 0.2 & 1.54463 & 0.673566 & 1.463 & 0.658 \\
0 & 0.5 & 1.7854  & 0.488086 & 1.469 & 0.616  \\
0 & 5   & 1.47941 & 0.868859 & 1.647 & 0.420 \\
0 & 10  & 1.800   & 0.410252 & 1.838 & 0.421 \\
0 & 20  & 2.15426 & 0.556954 & 2.154 & 0.544 \\ \\
1 & 0.1 & 1.93407 & 0.750046 & 2.021 & 0.652  \\
1 & 0.2 & 1.90391 & 0.773947 & 2.024 &  0.637 \\
1 & 0.5 & 1.94463 & 0.734591 & 2.035 &  0.602\\
1 & 5   & 2.44511 & 0.455166 & 2.348 & 0.434 \\
1 & 10  & 2.6644 & 0.463966 &  2.670 & 0.458 \\
1 & 20  & 3.2276 & 0.581024 & 3.209 & 0.590 \\ 
\end{tabular} 
\end{ruledtabular}
\end{table}

In the time domain the signal has three stages: the initial pulse
dependent on the source of perturbations, the quasinormal ringing
dominating period, and the power-law tail (see Fig.\ref{fig_GB1}).
The bigger GB-coupling is, the larger the quasinormal dominated region, 
i.e. at later times the tails start dominating. 
As can be seen from Fig.\ref{fig_GB2}, the power-law
tails do not show dependence on the Gauss-Bonnet coupling
$\alpha$ and are the same as for the d-dimensional Schwarzschild 
black hole in Einstein general relativity, when $d$ is odd. 
That is, the fields always shows a power-law falloff: 
for odd $d>3$ the field behaves as 
\begin{equation}
R_{\ell} \propto t^{-(2\ell + d - 2)} 
\end{equation}
at late times, where  $\ell$ is the multipole number. This behavior is
entirely due to $d$ being odd and  does not depend on the presence of
a black hole \cite{Vitor-Tail}. It is known, that in Einstein gravity,
for even $d>4$, the field decays as  \cite{Vitor-Tail}  
\begin{equation}
R_{\ell} \propto t^{-(2\ell + 3d - 8)} \quad ,
\end{equation}
and for the latter case there is no contribution from the flat background. 
This power-law tail is entirely due to the presence of the black hole
\cite{Vitor-Tail}. At the same time, the Gauss-Bonnet black hole
metric (\ref{metric_form}-\ref{metric}) goes to pure Minkowskian
metric when the black hole mass equals zero, i.e. in a space-time
without a black hole. In other words, empty space-time in Gauss-Bonnet
gravity ``does not see'' the $\alpha$. That is why we do not observe
the $\alpha$-dependence of tails in odd space-time dimensions.  

Thus, if the Gauss-Bonnet term changes late-time behavior, it must show itself
only for even-dimensional space-time. In the numerical procedure
developed with the characteristic integration scheme, no
tails (power-law or otherwise) were observed in even dimensions
Gauss-Bonnet spherical black holes. Yet, it should be pointed out that the
integration of the scalar field equation in the GB background is a
much more demanding numerical problem than the same integration with
the usual Einstein coupling. In the latter case there are auxiliary
analytical results, such as explicit expression for the tortoise
coordinate function. Therefore, the GB codes are less precise and more
time consuming, and eventual tails could be hidden. The possible
absence of tails with even $d$ deserves further consideration.

\subsection{Gauss-Bonnet-de Sitter black holes}

For the GBdS black holes the quasinormal ringing stage becomes correlated
with a new parameter: a positive $\Lambda$-term. When the $\Lambda$-term
is growing, both real oscillation frequency and the damping rate are
decreasing. Yet, real part of $\omega$ is more sensitive to the changes
of  $\Lambda$-term. 

\begin{figure}
\resizebox{1\linewidth}{!}{\includegraphics*{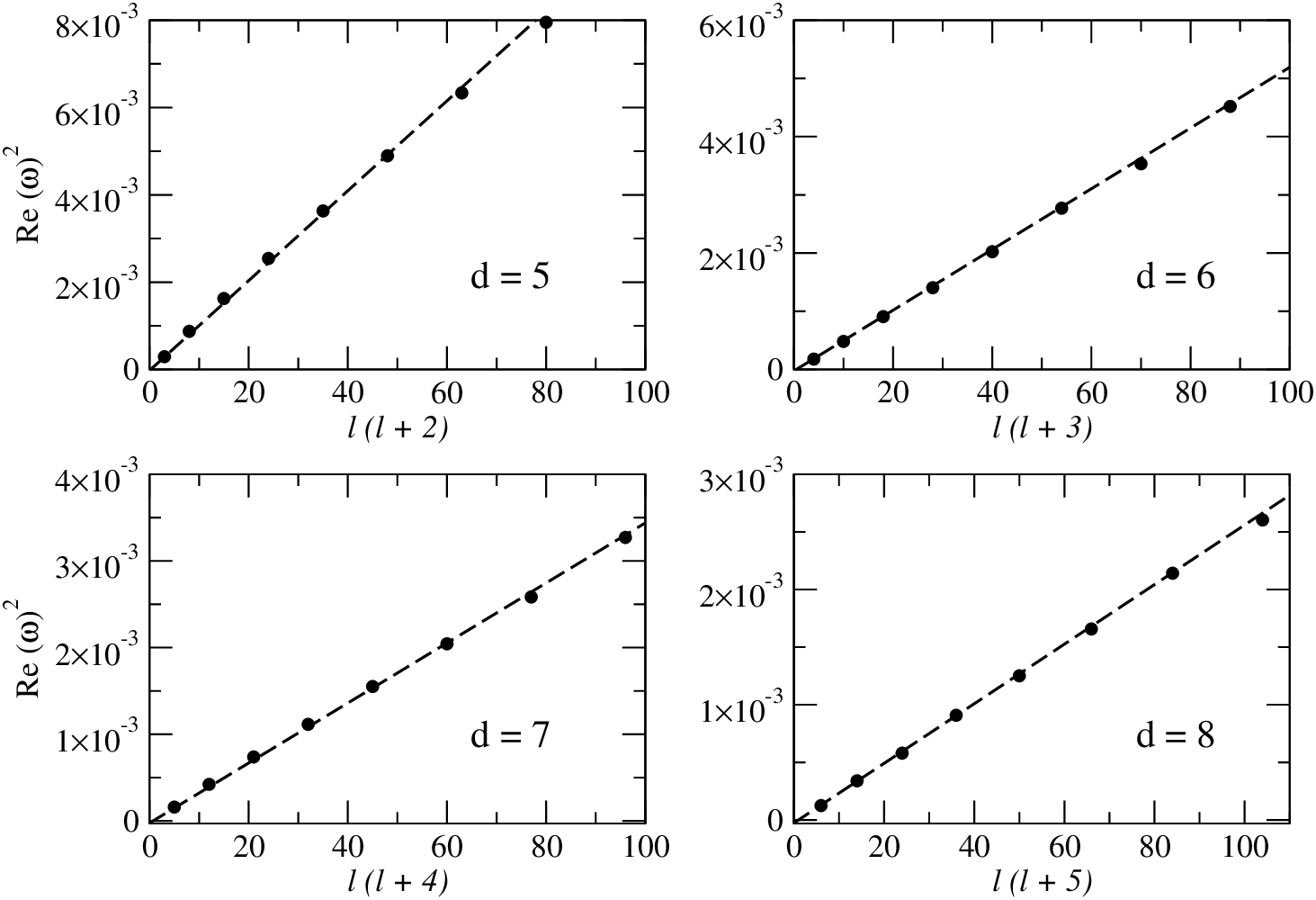}}
\caption{Graph of $\textrm{Re} (\omega)^2$ as a function of $\ell
    (\ell + d - 3)$, in the near-extreme limit positive $\Lambda$
    limit. The bullets are the values calculated from the
    time-evolution profiles, and the dashed lines are values obtained
    from the expression (\ref{QNM_near_extreme}). The parameters in
    this graph are $\kappa_{+}=0.01$, $\alpha=1$, $\mu=1.0$, and the
    differences between the analytical and numerical results are under
    2\%.} 
\label{near_extreme}
\end{figure}

Qualitatively this resembles the quasinormal
oscillations of $d$-dimensional  Schwarzschild-de Sitter black hole
\cite{konoplya03}. In the limit of extremal value of the
$\Lambda$-term, i.e. when the cosmological horizon ($r_{c}$) is very
close to the event horizon ($r_{+}$), it is possible to generalize the
formulas found in \cite{Cardoso-03} for four dimensional black hole
and in \cite{Molina} for $d$-dimensional case.  Namely, the
quasinormal frequencies for the near extreme Gauss-Bonnet
asymptotically de Sitter black holes are given by  
\begin{gather}
\frac{\omega_{n}}{\kappa_{+}} =  \sqrt{ \frac{\ell(\ell+d-3)}{r_{+}^{2}}
  \frac{r_{c} - r_{+}}{2\kappa_{+}} - \frac{1}{4}} -i
  \left(n+\frac{1}{2}\right)  \quad , \nonumber \\  
n = 0, 1, 2, \ldots \quad ,
\label{QNM_near_extreme}
\end{gather}
where $\kappa_{+}$ (a function of $\alpha$ in this generalized
context) is the surface gravity at the event horizon. The above
formula is well confirmed numerically: Fig.\ref{near_extreme} shows a
comparison of the values obtained by direct numerical calculation and
from Eq. (\ref{QNM_near_extreme}). Thus the quasinormal modes are
proportional to the surface gravity $\kappa_{+}$, at least for lower
overtones. It should be pointed that for the usual Schwarzschild-de
Sitter black holes, numerical and analytical investigations
\cite{SdShigh} suggest that the high overtone behavior does not obey
the formula (\ref{QNM_near_extreme}).

%%%%%%%%%%%%%%%%%%%%%%%%%%%%%%%%%%%%%%%%%%%%%%%%%%%%%%%%%%%%
%%%%%%%%%%%%%%%%%%%%%%%%%%%%%%%%%%%%%%%%%%%%%%%%%%%%%%%%%%%%

When the mass parameter $\mu$ is set to zero, we have the case of pure
de Sitter spacetime in the Gauss-Bonnet gravity. The metric function
(\ref{metric}) then reduces to the following form:
\begin{equation}
h(r) = 1 + \frac{r^2}{2\alpha} - \frac{r^2}{2\alpha} 
       \sqrt{1 + \frac{8\alpha \Lambda}{(d-1)(d-2)} } \quad .
\end{equation}
Repeating the analysis of \cite{Abdalla-02,natario}, we
come to the conclusion that quasinormal modes exist only in odd
spacetime dimensions and are given by the formula:
\begin{gather}
\omega_{n} = i\left[\frac{1} {2\alpha}\left(1 - \sqrt{1 + \frac{8\alpha
\Lambda}{(d-1)(d-2)}}\right)\right]^{1/2} (2 n + \ell) \nonumber \\
n = 0, 1, 2, \ldots \quad .
\end{gather}
Note that pure Gauss-Bonnet-de Sitter quasinormal modes are purely
imaginary, which corresponds to exponential decaying without
oscillations.

\begin{figure}
\resizebox{1\linewidth}{!}{\includegraphics*{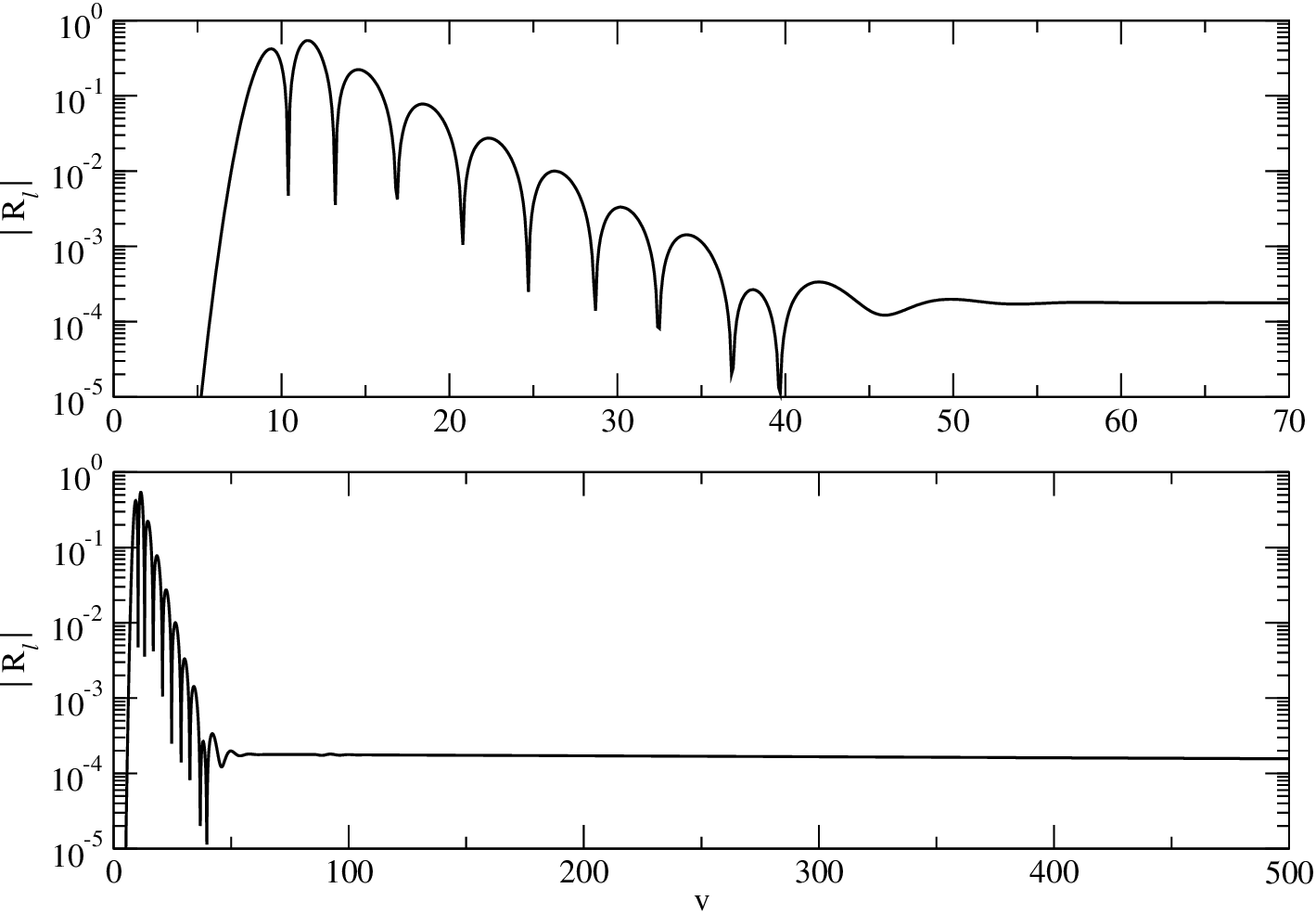}}
\caption{Field decay in the Gauss-Bonnet-de Sitter
    black holes, with $\ell=0$.  A quasinormal mode
    dominated region is observed (above), and asymptotically the field
    decays to a constant (below). The fundamental mode, calculated with the WKB
    method and directly from the characteristic data are $0.8356 - 0.2935i$ and
    $0.8011 - 0.2608i$. The parameters in this graph are $d=6$, $\alpha=3.0$,
    $\Lambda=0.1$ and $\mu=1.0$.}
\label{fig_GBdS1}
\end{figure}

It is well-known  that the late-time tails of black holes in
asymptotically de Sitter space-time for
zero multipole and for higher multipoles are qualitatively different.
For the zero multipole field ($\ell=0$), the time domain picture is
the following: after a transient part, a
quasinormal mode dominated region is best observed. Following the
quasinormal mode dominated region, a late-time decay region settles.
In this latter phase, the  wave-functions decay asymptotically to a
constant value, as has been the case in the Schwarzschild de Sitter
black hole which was studied before
\cite{Brady-97,Brady-99,Molina-04}. 
This is illustrated in Fig.\ref{fig_GBdS1}.  

\begin{figure}
\resizebox{1\linewidth}{!}{\includegraphics*{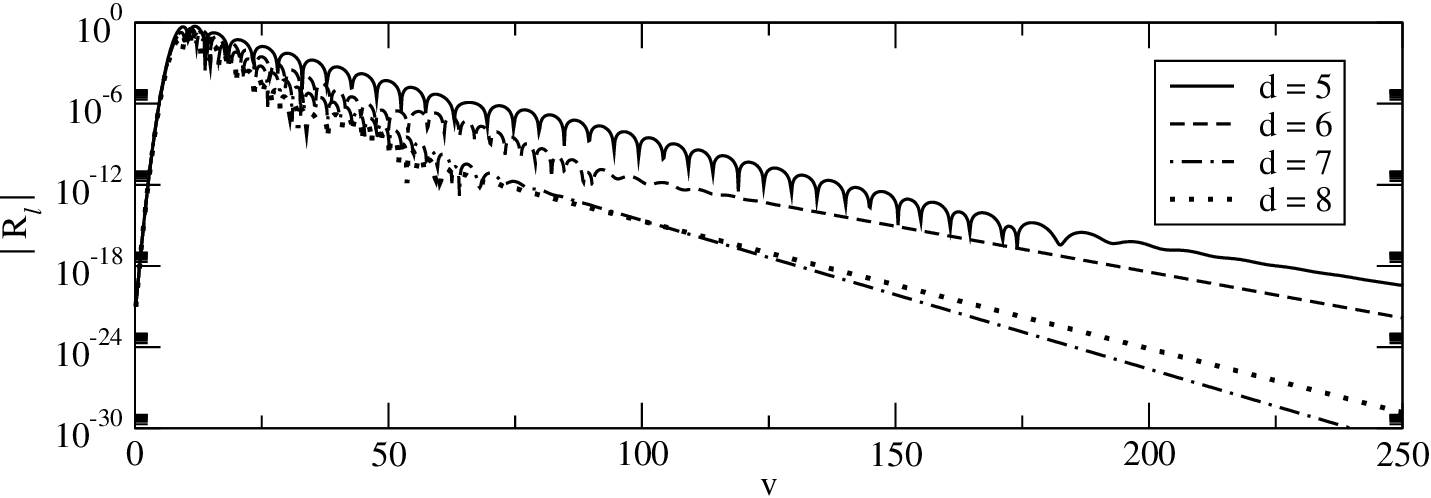}}
\caption{Field decay in the Gauss-Bonnet-de Sitter 
    black holes, for $d=5,6,7,8$. It is observed a quasinormal mode
    dominated region. Asymptotically, the field decays as an
    exponential tail.  The parameters in this graph are $\alpha=0.1$,
    $\mu=1.0$ and $\ell=1$.}
\label{fig_GBdS2}
\end{figure}

For first and higher multipoles ($\ell>0$) at late times we observe
exponential tails in vicinity of GBdS black hole. This is also an
expected result, since in Einstein gravity the exponential tails are
observed as well in usual de Sitter black holes
\cite{Brady-97,Brady-99,Molina-04}.  This is illustrated in
Fig.\ref{fig_GBdS2}. The dependence of the quasinormal modes 
on $\Lambda$-term and Gauss-Bonnet coupling can be learnt 
from Tables X-XII for different space-time dimensionality.

%%%%%%%%%%%%%%%%%%%%%%%%%%%%%%%%%%%%%%%%%%%%%%%%%%%%%%%%%%%%
%%%%%%%%%%%%%%%%%%%%%%%%%%%%%%%%%%%%%%%%%%%%%%%%%%%%%%%%%%%%

%%%%%%%%%%%%%%%%%%%%%%%%%%%%%%%%%%%%%%%%%%%%%%%%%%%%%%%%%%%%
%%%%%%%%%%%%%%%%%%%%%%%%%%%%%%%%%%%%%%%%%%%%%%%%%%%%%%%%%%%%

The numerical simulations developed for Gauss-Bonnet-de Sitter
black hole indicate that the massless scalar perturbation in
this geometry behave asymptotically as 
\begin{eqnarray}
R_{\ell}\approx \exp \left[ \ell \left(\kappa_{c} + c^{quad}
\kappa_{c}^{2}\right) t \right] & \textrm{as} & t\rightarrow\infty
\quad , \nonumber \\ 
d \ge 4  \quad \textrm{and} \quad \alpha \ge 0  \quad ,
\label{tails-GBdS}
\end{eqnarray}
where $\kappa_{c}$ is the surface gravity at the cosmological
horizon and $c^{quad}$ is an adjustment parameter for the
$\kappa_{c}^{2}$ correction. The expression (\ref{tails-GBdS}) shows
that, although the exponential tail in the GBdS background is
dependent on the Gauss-Bonnet coupling $\alpha$ (since $\kappa_{c}$ is
a function of $\alpha$), the form of the dependence is identical to
the null $\alpha$ case. Eq. (\ref{tails-GBdS}) generalizes the
analogous expression found in \cite{Molina-04} for the usual
Schwarzschild black holes.

\begin{table}
\caption{Values for the quasinormal frequencies for the fundamental
  mode in the Gauss-Bonnet-de Sitter geometry, obtained from third and
  sixth order WKB method, for
  $d = 5$, $\ell=0$ and several values of $\alpha$ and $\Lambda$.} 
\label{GB-dS-L0-1} 
\begin{ruledtabular}
\begin{tabular}{cccccc}
\multicolumn{2}{l}{ d = 5}         &
\multicolumn{2}{c}{WKB (3th order)}&
\multicolumn{2}{c}{WKB (6th order)}\\ 
$\alpha$ & $\Lambda$ & Re($\omega_0$) & -Im($\omega_0$)  &
Re($\omega_0$) & -Im($\omega_0$)
\\
\hline
\\
0.1 & 1/8 & 0.30485   & 0.278407 & 0.334981   & 0.26202 \\
0.1 & 1/2 & 0.14922   & 0.217542 & 0.152444   & 0.2164  \\ 
0.1 & 2/3 & 0.0677006 & 0.14792  & 0.00666957 & 0.150988  \\ \\
1   & 1/5 & 0.301869  & 0.218752  & 0.378865  & 0.159611 \\
1   & 1   & 0.112613  & 0.148749  & 0.11619   & 0.145465 \\ 
1   & 7/5 & 0.0240051 & 0.0683652 & 0.0242716 & 0.0709775  \\
\end{tabular} 
\end{ruledtabular}
\end{table}

\begin{table}
\caption{Values for the quasinormal frequencies for the fundamental
  mode in the Gauss-Bonnet-de Sitter geometry, obtained from third and
  sixth order WKB method, for $d = 6$, $\ell=0$ and several values of
  $\alpha$ and $\Lambda$.}  
\label{GB-dS-L0-2} 
\begin{ruledtabular}
\begin{tabular}{cccccc}
\multicolumn{2}{l}{ d = 6}         &
\multicolumn{2}{c}{WKB (3th order)}&
\multicolumn{2}{c}{WKB (6th order)}\\ 
$\alpha$ & $\Lambda$ & Re($\omega_0$) & -Im($\omega_0$)  &
Re($\omega_0$) & -Im($\omega_0$)
\\
\hline
\\
0.1 & 1/4 & 0.580472   & 0.428367  & 0.652079  & 0.411337 \\
0.1 & 1   & 0.379931   & 0.392027  & 0.401925  & 0.384688 \\ 
0.1 & 2   & 0.0653851  & 0.168544  & 0.0636328 & 0.173794  \\ \\
1   & 1/2 & 0.582293   & 0.358225  & 0.736989  & 0.235326 \\
1   & 2   & 0.318697   & 0.307084  & 0.331268  & 0.282461  \\
1   & 4   & 0.00390515 & 0.0487306 & 0.0574555 & 0.0529852 \\ \\
10  & 100 & 1.5033     & 0.574511  & 1.5741    & 0.599309 \\
\end{tabular} 
\end{ruledtabular}
\end{table}

\begin{table}
\caption{Values for the quasinormal frequencies for the fundamental
  mode in the Gauss-Bonnet-de Sitter geometry, obtained from third and
  sixth order WKB method, for $d = 7$, $\ell=0$ and several values of
  $\alpha$ and $\Lambda$.}  
\label{GB-dS-L0-3} 
\begin{ruledtabular}
\begin{tabular}{cccccc}
\multicolumn{2}{l}{ d = 7}         &
\multicolumn{2}{c}{WKB (3th order)}&
\multicolumn{2}{c}{WKB (6th order)}\\ 
$\alpha$ & $\Lambda$ & Re($\omega_0$) & -Im($\omega_0$)  &
Re($\omega_0$) & -Im($\omega_0$)
\\
\hline
\\
0.1 & 1 & 0.769109   & 0.556647 & 0.852395  & 0.551612 \\
0.1 & 1 & 0.561644   & 0.517639 & 0.598484  & 0.511069 \\ 
0.1 & 1 & 0.00963007 & 0.222564 & 0.0939457 & 0.228191  \\ \\
1   & 1 & 0.868246   & 0.48557  & 1.12603   & 0.2999562 \\
1   & 4 & 0.461426   & 0.419016 & 0.483098  & 0.381285  \\
1   & 7 & 0.113334   & 0.215329 & 0.111663  & 0.218394 \\
\end{tabular} 
\end{ruledtabular}
\end{table}

\begin{table}
\caption{Values for the quasinormal frequencies for the fundamental
  mode in the Gauss-Bonnet-de Sitter geometry, obtained from third and
  sixth order WKB method, for $d = 8$, $\ell=0$ and several values of
  $\alpha$ and $\Lambda$.}  
\label{GB-dS-L0-4} 
\begin{ruledtabular}
\begin{tabular}{cccccc}
\multicolumn{2}{l}{ d = 8}         &
\multicolumn{2}{c}{WKB (3th order)}&
\multicolumn{2}{c}{WKB (6th order)}\\ 
$\alpha$ & $\Lambda$ & Re($\omega_0$) & -Im($\omega_0$)  &
Re($\omega_0$) & -Im($\omega_0$)
\\
\hline
\\
0.1 & 1 & 1.1449   & 0.669364 & 1.29739  & 0.695376 \\
0.1 & 4 & 0.644892 & 0.598472 & 0.682218 & 0.594199 \\
\end{tabular} 
\end{ruledtabular}
\end{table}

\begin{table}
\caption{Values for the quasinormal frequencies for the fundamental
  mode in the Gauss-Bonnet-de Sitter geometry, obtained from third and
  sixth order WKB method and directly from characteristic data, for
  $d = 5$, $\ell=1$ and several values of $\alpha$ and $\Lambda$.} 
\label{GBdS_L1_1}
\begin{ruledtabular}
\begin{tabular}{cccccc}
\multicolumn{2}{l}{ d = 5 }        &
\multicolumn{2}{c}{WKB (6th order)}&
\multicolumn{2}{c}{Characteristic Integration}\\
  $\alpha$ & $\Lambda$ & 
  Re($\omega_0$) & -Im($\omega_0$) & 
  Re($\omega_0$) & -Im($\omega_0$) \\ 
\hline \\
0.1 & 1/8  & 0.64317  & 0.241723 & 0.6451 & 0.2382 \\
0.1 & 1/2  & 0.379265 & 0.165101 & 0.3856 & 0.1564 \\ 
0.1 & 2/3  & 0.231024 & 0.103394 & 0.2247 & 0.1051 \\ \\
1 & 1/5    & 0.698655 & 0.181854  & 0.6926 & 0.1877  \\
1 & 1     & 0.365998 & 0.111668  & 0.3704 & 0.1133 \\ 
1 & 7/5  & 0.153506 & 0.0465747 & 0.1518 & -0.042631 \\ 
\end{tabular} 
\end{ruledtabular}
\end{table}

\begin{table}
\caption{Values for the quasinormal frequencies for the fundamental
  mode in the Gauss-Bonnet-de Sitter geometry, obtained from third and
  sixth order WKB method and directly from characteristic data, for
  $d = 6$, $\ell=1$ and several values of  $\alpha$ and $\Lambda$.} 
\label{GBdS_L1_2}
\begin{ruledtabular}
\begin{tabular}{cccccc}
\multicolumn{2}{l}{ d = 6 }        &
\multicolumn{2}{c}{WKB (6th order)}&
\multicolumn{2}{c}{Characteristic Integration}\\
  $\alpha$ & $\Lambda$ & 
  Re($\omega_0$) & -Im($\omega_0$) & 
  Re($\omega_0$) & -Im($\omega_0$) \\
\hline \\
0.1 & 1/4  & 1.04698  & 0.39859  & 1.0574 & 0.3828  \\
0.1 & 1    & 0.74644  & 0.323514 & 0.7479 & 0.3183  \\ 
0.1 & 2    & 0.238066 & 0.114496 & 0.2387 & 0.1212  \\ \\
1 & 1/2   & 1.09756   & 0.301244  & 1.053   & 0.2948 \\
1 & 2     & 0.6962    & 0.244326  & 0.6962  & 0.2307 \\ 
1 & 4 & 0.0847573 &0.0315107&0.08965&0.03323 \\ \\
10 & 100  & 3.21004 & 0.52559 & 3.1940 & 0.5360 \\
10 & 2000         & -       & -       & 2.1711  & 0.4900 \\ 
10 & 5000         & -       & -       & 0.6228 & 0.2018 \\   
\end{tabular} 
\end{ruledtabular}
\end{table}

\begin{table}
\caption{Values for the quasinormal frequencies for the fundamental
  mode in the Gauss-Bonnet-de Sitter geometry, obtained from third and
  sixth order WKB method and directly from characteristic data, for
  $d = 7$, $\ell=1$, $\alpha=0.1$ and several values of $\Lambda$.} 
\label{GBdS_L1_3}
\begin{ruledtabular}
\begin{tabular}{cccccc}
\multicolumn{2}{l}{ d = 7 }        &
\multicolumn{2}{c}{WKB (6th order)}&
\multicolumn{2}{c}{Characteristic Integration}\\
  $\alpha$ & $\Lambda$ & 
  Re($\omega_0$) & -Im($\omega_0$) & 
  Re($\omega_0$) & -Im($\omega_0$) \\
\hline \\
0.1 & 1  & 1.28621  & 0.519229 & 1.304  & 0.4910  \\
0.1 & 2  & 1.00125  & 0.440876 & 1.008  & 0.4306 \\ 
0.1 & 4  & 0.303405 & 0.153651 & 0.3091 & 0.1416 \\ \\
1 & 1  & 1.48546  & 0.411346 & 1.467  & 0.3623 \\
1 & 4  & 0.899964 & 0.338928 & 0.8944 & 0.3224 \\ 
1 & 7  & 0.361602 & 0.152214 & 0.3638 &0.1398 \\ 
\end{tabular} 
\end{ruledtabular}
\end{table}

\begin{table}
\caption{Values for the quasinormal frequencies for the fundamental
  mode in the Gauss-Bonnet-de Sitter geometry, obtained from third and
  sixth order WKB method and directly from characteristic data, for
  $d = 8$, $\ell=1$, $\alpha=0.1$ and several values of $\Lambda$.} 
\label{GBdS_L1_4}
\begin{ruledtabular}
\begin{tabular}{cccccc}
\multicolumn{2}{l}{ d = 8 }        &
\multicolumn{2}{c}{WKB (6th order)}&
\multicolumn{2}{c}{Characteristic Integration}\\
  $\alpha$ & $\Lambda$ & 
  Re($\omega_0$) & -Im($\omega_0$) & 
  Re($\omega_0$) & -Im($\omega_0$) \\
\hline \\
0.1 & 1 & 1.75036 & 0.697491 & 1.803 & 0.6250 \\
0.1 & 4 & 1.10954 & 0.508301 & 1.103 & 0.4964 \\ 
\end{tabular} 
\end{ruledtabular}
\end{table}

\subsection{Gauss-Bonnet-anti-de Sitter black holes}    

The quasinormal and late-time behavior of black holes 
in anti-de Sitter spacetime is significantly different from those 
in asymptotically de Sitter or flat spacetimes. The key difference is
stipulated by the effective potential behavior, which is divergent at
spacial infinity. Thus the anti-de Sitter space acts as an effective
confining box. Therefore the Dirichlet boundary conditions are
natural. These boundary conditions  are required also by AdS/CFT
correspondence for scalar field perturbations \cite{Horowitz00}. Yet, for 
higher spin perturbations the true boundary conditions may be
different \cite{moss}.  
 
In the usual Schwarzschild-anti-de Sitter black holes,
the quasinormal modes govern the decay at all times and thereby no power-law or
exponential tails appear \cite{Wang-01,Wang-04}. We observe a similar
behavior the scalar field perturbations in the Gauss-Bonnet-anti-de
Sitter black holes.  

\begin{figure}
\resizebox{1\linewidth}{!}{\includegraphics*{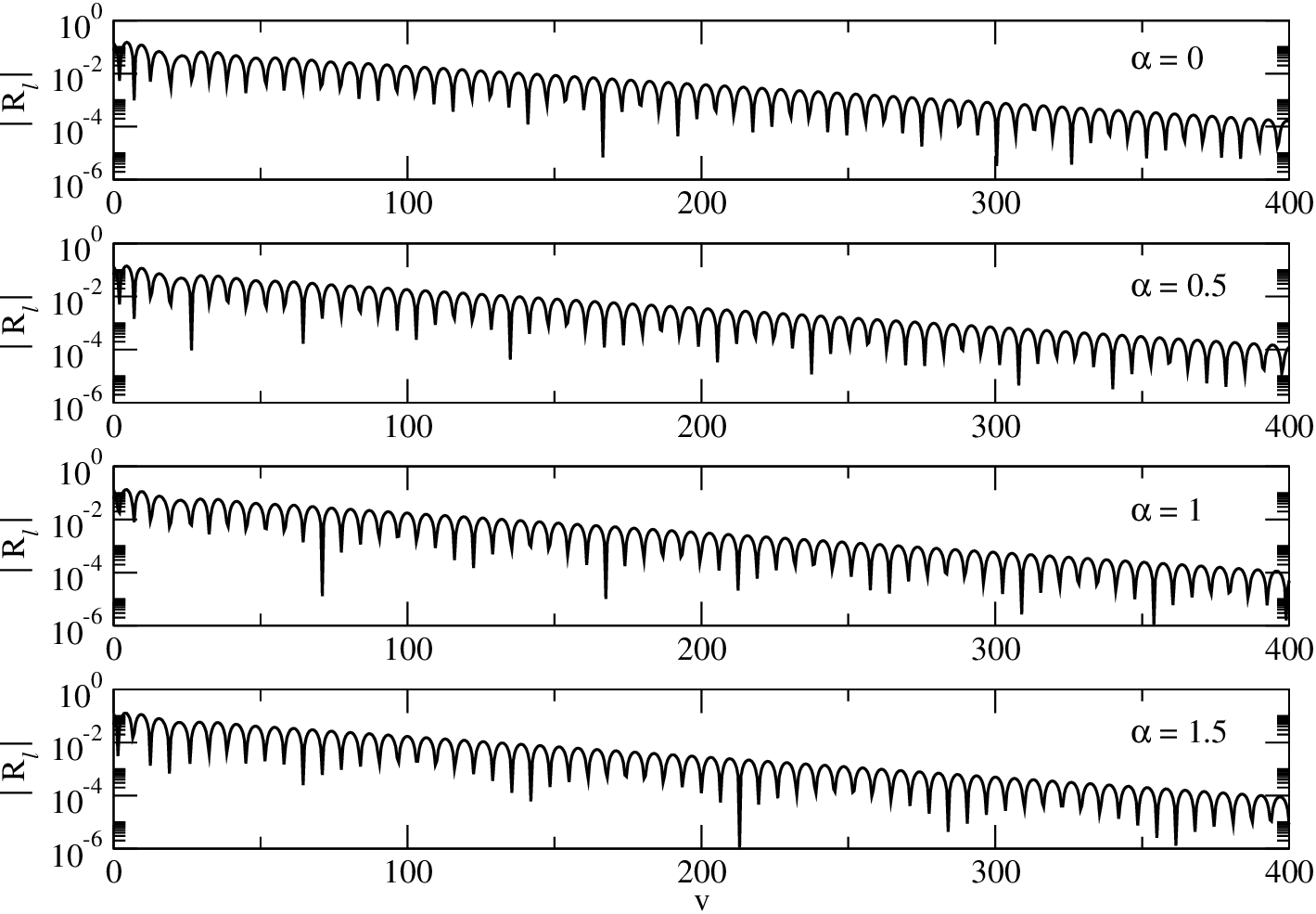}}
\caption{Field decay in the Gauss-Bonnet anti-de Sitter
    black holes, for several values of the Gauss-Bonnet coupling. It
    is observed a quasinormal mode dominated region. Asymptotically,
    the field decays in  quasinormal modes. The parameters in this
    graph are $d=5$, $\mu=1.0$ $\Lambda= -0.1$ and $\ell=0$.}
\label{fig_GBAdS1}
\end{figure}

It is not possible to use WKB method to find the quasinormal modes in
Gauss-Bonnet-AdS case because the effective potential is not a
potential barrier anymore. The Horowitz-Hubeny method
\cite{Horowitz00} is not applicable either, because the Taylor expansion
of the effective potential has infinite number of terms. That is why
we were limited only by time domain analysis, which is free of the
above problems. From Fig.\ref{fig_GBAdS1} we see that, indeed, the
quasinormal modes are dominating even at sufficiently late times. We
also have learnt  from Fig.\ref{fig_GBAdS1}  that  the
quasinormal mode dominated region grows, as the  multipole index
$\ell$ grows.       

As is known from Einstein action case, as the radius of the AdS black
hole goes to zero, the quasinormal modes of the black hole approach
its pure anti-de Sitter values \cite{konoplya02}. Repeating the
calculations of \cite{natario}, we find the exact
expression for the normal modes in GB gravity: 
\begin{gather}
\omega_{n} = \left[ \frac{1} {2\alpha}\left(1 - 
\sqrt{1 + \frac{8\alpha \Lambda}{(d-1)(d-2)}}\right)\right]^{1/2} (2 n +
\nonumber \\
\ell + d - 1) \quad ,
\nonumber \\
n = 0, 1, 2, \ldots \quad .
\end{gather}
The pure GB-AdS modes, unlike GB-dS modes, exist in any any spacetime
dimension.

\begin{table}
\caption{Values for the quasinormal frequencies for the fundamental
  mode in the Gauss-Bonnet-anti-de Sitter geometry, estimated from the
  characteristic data, for $d = 5$, $\ell=0$, $\mu = 1.0$, $\Lambda=
  -0.1$ and several values of $\alpha$.}  
\label{GBAdS_L01_d5}
\begin{ruledtabular}
\begin{tabular}{ccc}
  $\alpha$ & 
  Re($\omega_0$) & -Im($\omega_0$) \\ 
\hline \\
0.1 & 0.4923 & -0.01585 \\
0.1 & 0.4920 & -0.01593 \\
0.5 & 0.4904 & -0.01634 \\ 
1.0 & 0.4885 & -0.01702 \\ 
1.5 & 0.4866 & -0.01766 \\ 
\end{tabular} 
\end{ruledtabular}
\end{table}

%%%%%%%%%%%%%%%%%%%%%%%%%%%%%%%%%%%%%%%%%%%%%%%%%%%%%%%%%%%%
%%%%%%%%%%%%%%%%%%%%%%%%%%%%%%%%%%%%%%%%%%%%%%%%%%%%%%%%%%%%

%%%%%%%%%%%%%%%%%%%%%%%%%%%%%%%%%%%%%%%%%%%%%%%%%%%%%%%%%%%%%%%%%%%%%%%%%%%%%%%
\section{Conclusions}
%%%%%%%%%%%%%%%%%%%%%%%%%%%%%%%%%%%%%%%%%%%%%%%%%%%%%%%%%%%%%%%%%%%%%%%%%%%%%%%

We have considered here frequency and time domain description of
evolution of scalar field perturbations in the exterior of black holes in
Gauss-Bonnet theory of gravity, generally with a $\Lambda$-term. The
quasinormal behavior even though being corrected by
a new parameter, Gauss-Bonnet coupling $\alpha$, are qualitatively
dependent mainly on the $\Lambda$-term and black hole parameters such
as mass $\mu$ and  multipole number $\ell$.  The late-time tails 
for asymptotically flat Gauss-Bonnet black holes, 
do not depend on the Gauss-Bonnet coupling in odd space-time
dimensions, and therefore are the same
as those for d-dimensional Schwarzschild black hole in Einstein
gravity. Moreover, in the case of Gauss-Bonnet-de Sitter black holes, 
the late-time tails, though dependent on $\alpha$, yet, rather
trivially, i.e. only through dependence of the surface gravity at the
cosmological radius on $\alpha$. Thus, the Gauss-Bonnet coupling shows
itself ``minimally'' in late-time behavior. The most interesting 
problem  which remains unsolved is, to find late time tails for 
even dimensions, and thereby, to know whether the power-law tails depend
upon the Gauss-Bonnet term.   
At the same time, we have shown that corrections to the quasinormal
frequencies due to GB-term are not negligible: they may reach $20\%$
for string theory predicted values of $\alpha \approx 1$.

Even though our analysis can easily be extended to the massive scalar
field, we were limited here by the massless case. We expect that the
influence of the massive term upon the QNMs will be similar to that
found in \cite{konoplyaPLB}, i.e. the lower overtones should be
corrected by the field mass, infinitely high overtone asymptotic will
be unchanged no matter the value of the massive term. 
Also we did not consider the high overtone behavior of the GB black holes.
Generally, the high overtone asymptotics must be studied by totally
different methods \cite{high-n} and deserves separate investigation.   

%%%%%%%%%%%%%%%%%%%%%%%%%%%%%%%%%%%%%%%%%%%%%%%%%%%%%%%%%%%%%%%%%%%%%%%%%%%%%%%

\begin{acknowledgments}
 This work was partially supported by \emph{Funda\c{c}\~{a}o de Amparo
\`{a} Pesquisa do Estado de S\~{a}o Paulo (FAPESP)} and \emph{Conselho
 Nacional de Desenvolvimento Cient\'ifico e Tecnol\'ogico (CNPq)}, Brazil.
\end{acknowledgments}

%%%%%%%%%%%%%%%%%%%%%%%%%%%%%%%%%%%%%%%%%%%%%%%%%%%%%%%%%%%%%%%%%%%%%%%%%%%%%%%

\end{document}